\documentclass{aastex}			
\usepackage{spr-astr-addons}	
\usepackage{url}

\usepackage{amsmath}			
\usepackage{textcomp}			
\usepackage{subfigure}			
\usepackage{enumitem}			
\usepackage{bm}					
\usepackage{xcolor}				
\usepackage{tabu}				
\usepackage{verbatim}
\usepackage{esdiff}				
\usepackage{soul}				

\newcommand{\E}{\hspace{2pt}\textsc{e}\hspace{1pt}}
\newcommand{\2}{\textonehalf}
\newcommand{\kB}{k_{\text{B}}}				
\newcommand{\GF}{G_\text{F}}				
\newcommand{\Gw}{G_\text{wk}}				
\newcommand{\GN}{G} 
\newcommand{\W}{$\sin^2 \theta_\text{W}$ }	
\newcommand{\ek}{e_{\text{k}}}				

\begin{document}
\title{Viscosity in a Lepton-Photon Universe}

\shorttitle{Viscosity in a Lepton-Photon Universe}
\shortauthors{Lars Husdal}

\author{Lars Husdal\altaffilmark{}}
\affil{Department of Physics, Norwegian University of Science and Technology,
	   Trondheim,
	   Norway}
\email{lars.husdal@ntnu.no}
\altaffiltext{1}{lars.husdal@ntnu.no}

\abstract{We look at viscosity production in a universe consisting purely of leptons and photons. This is quite close to what the universe actually look like when the temperature was between $10^{10}$ K  and $10^{12}$ K ($1$ -- $100$ MeV). By taking  the strong force and the hadronic particles out of the equation, we can examine how the viscous forces behave with all the 12 leptons present. By this we study how shear- and (more interestingly) bulk viscosity is affected during periods with particle annihilation. We use the theory given by Hoogeveen et. al. from 1986, replicate their 9-particle results and expand it to include the muon and tau particles as well. This will impact the bulk viscosity immensely for high temperatures. We will show that during the beginning of the lepton era, when the temperature is around 100 MeV, the bulk viscosity will be roughly 100 million times larger with muons included in the model compared to a model without.}


\section{Introduction}
\label{Sec:Introduction}

\begin{figure}[t]
	\centering
	\includegraphics[width=\columnwidth]{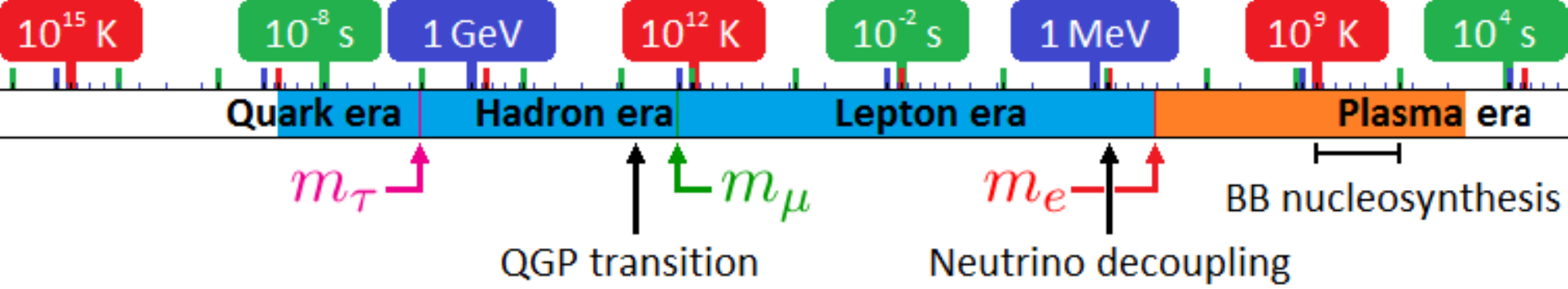}
	\caption{The temperature (in kelvin and MeV), and time scales used in our paper.}
	\label{Fig:Timeline}
\end{figure}

Viscosity, the resistance to gradual deformation, is a well-described phenomenon in classical fluid dynamics. Shear (or dynamic) viscosity is the resistance to shearing flows, and occurs when adjacent layers of a fluid, which are parallel to each other, move at different velocities, or have different temperatures. Bulk (or volume) viscosity is the internal resistance for a fluid to evenly expand (or compress). Both types of viscosity played an important role in the early universe. See Fig. \ref{Fig:Timeline}.

Several papers about viscous cosmology have been published over the years \citep{Hoogeveen:1986, Treciokas:1971, Gron:1990, Brevik:2015}. Most of them have looked upon it from a phenomenological point of view.

But what does this all mean in a cosmological framework? Or more specifically during the lepton era? At this time the universe consisted of photons, charged leptons and neutrinos --- more or less acting as ideal (relativistic and non-relativistic) gases. The early universe was also very close to being in local (and global) thermal equilibrium, it was almost completely isotropic.

\begin{figure}[h]
	\centering
	\includegraphics[width=\columnwidth, height=4cm]{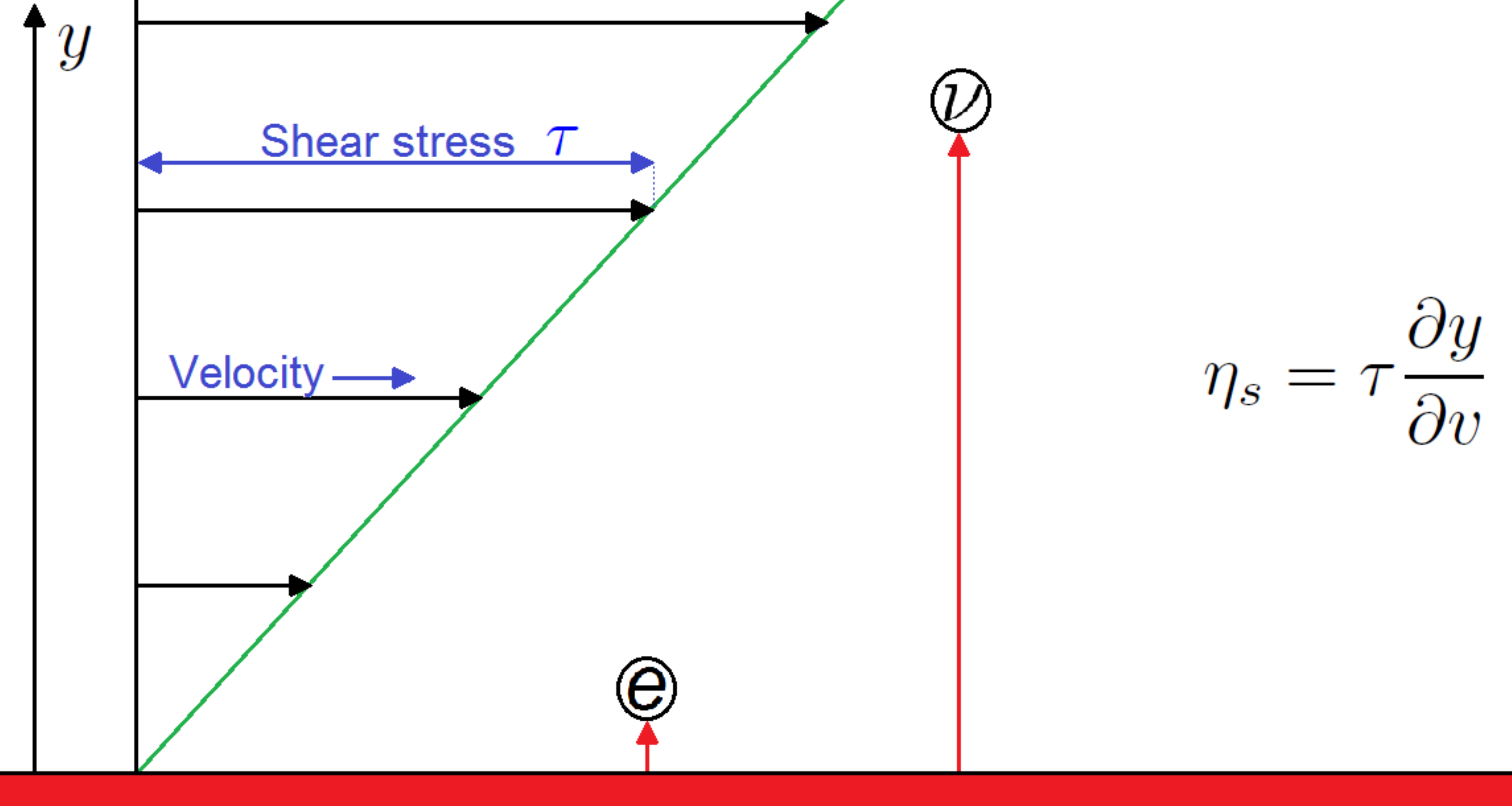}
	\caption{A simple illustration of shear viscosity. The weakly interaction neutrinos have a much longer mean free path than the electromagnetic interacting particles (e.g. like electrons). They will thus travel much farther in the perpendicular direction ($y$) of the flow.}
	\label{Fig:ShearViscosityIllustration}
\end{figure}

Let us give a simple qualitative description of how the two previously mentioned types of viscosities acted in this era: Any stress forces driving the system towards anisotropy is countered and resisted by viscous forces. In the first case we have shear stress due to regional differences in flow or temperature. The way shear viscosity drives the system back towards equilibrium is through momentum transfer between the regions. See Fig. \ref{Fig:ShearViscosityIllustration}. This is proportional to the mean free path of the momentum-carrying particles. Because the weak nuclear force is so much weaker than the electromagnetic force at this time, the mean free path of neutrinos are much larger than those of the electromagnetic interacting particles. Practically, all the action of shear viscosity is due the interactions between neutrinos and other particles. 

\begin{figure}[h]
	\centering
	\includegraphics[width=\columnwidth]{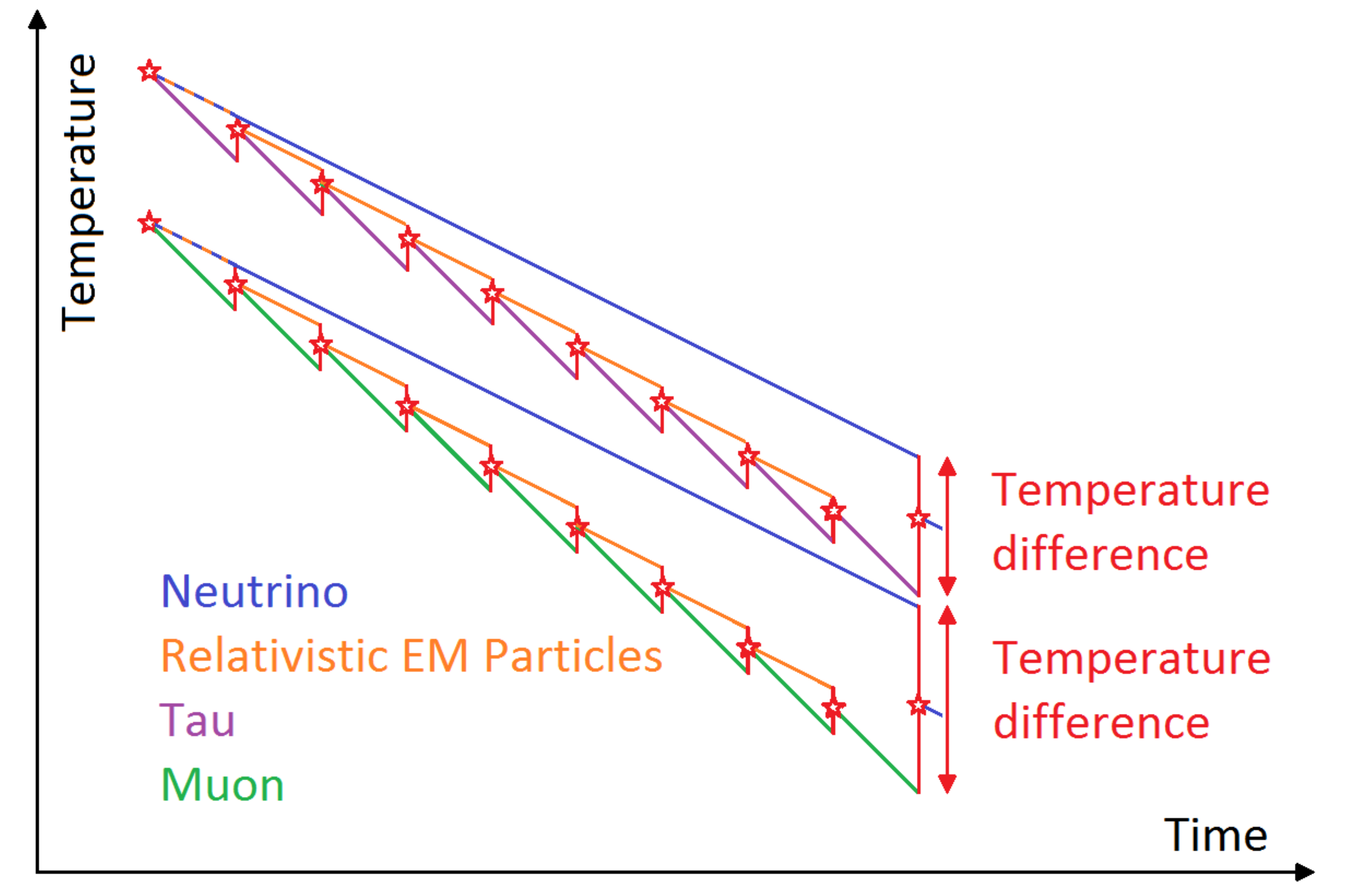}
	\caption{A simplified illustrating of bulk viscosity. The electromagnetic interacting particles ($\gamma, e, \mu, \tau$) all have a short mean life time, resulting in a small momentum transfer. The longer lived neutrinos provide a much larger momentum transfer since the temperature difference is bigger. The heavier leptons (here illustrated by the purple $\tau$ and $ \bar{\tau}$ particles) cool down slower compared to the a lighter lepton (here illustrated by the green $\mu$ and $ \bar{\mu}$ particles), since	they are heated up by more relativistic particles.}
	\label{Fig:BulkViscosityIllustration}
\end{figure}

Bulk viscosity arises due to the fact that non-relativistic and relativistic gases behave adiabatically differently. During expansion, relativistic gases drop in temperature as $a^{-1}$, while non-relativistic gases drop as $a^{-2}$ (where $a$ is the scale factor). The system as a whole will try to bring these two components into equilibrium by heat transfer. Also here the main component of the momentum transfer is due to the neutrinos, but is a bit more complicated. As the temperature of the universe cools, and some particles  transitions towards non-relativistic velocities they will be start to drop in temperature according as the aforementioned $a^{-2}$. They will be heated up by all the relativistic particles (which themselves will be cooled down). The electromagnetic interacting particles will never have time to gain any real temperature difference and they collectively will cool down a bit faster. The neutrinos on the other hand, because of their weakly interacting nature, have a longer mean life time resulting in a larger temperature difference before they interact. This will lead to a larger momentum transfer and a resulting larger viscous effect. In this way the photons play an indirect role as they help heat up the non-relativistic particles such that the temperature difference between them and the neutrinos are smaller than it would be the case in a non-photon case. See Fig. \ref{Fig:BulkViscosityIllustration}. From the same argument it follows that compared to the neutrinos, the muons should drop faster in temperature than the tau-leptons, and the electrons even faster. We will later see that bulk viscosity also has another impact as it is associated with a pressure.

Being interested in the underlying physics, we looked at some earlier papers with a focus on the origins for shear and bulk viscosity, including  \cite{Misner:1967, Misner:1968, Misner:1969}, \cite{Caderni:1977}, \cite{vanErkelens:1978} \cite{vanLeeuwen:1986} and \cite{Hoogeveen:1986} with a special focus on the latter (``Viscous Phenomena in Cosmology. I. Lepton Era").  We used their model, reproduced their result and took a closer look at the transitions temperatures towards the colder plasma era and the warmer hadron and quark eras. We were interested in the evolution of the two viscosity coefficients.

The main purpose with this work is to study the impact of including all three lepton generations. As these will annihilate at different times, we can study the build-up and fall-down of the two viscosity coefficients. We will also look at some aspects where this model comes short.

\section{Setup and constraints}
\label{Sec:Constraints}
\subsection{Definition of lepton era}
The lepton era (or epoch) follows not long after the end of the phase transition from quark-gluon plasma to hadron gas at roughly $170~\text{MeV}$ ($1.9 \times 10^{12}$ K) \citep{Kapusta:QGP}. At this time, the free quarks and gluons bound together to form mesons and baryons. Because the temperature is too low for most pair productions, the more massive baryons quickly die out, and soon only the lightest mesons --- the pions, have a significant contribution to the particle-mix. Already at the time of this phase transition, the number of leptons (e, $\mu$, $\tau$, $\nu$) equals that of the hadrons, and the energy density is soon after ($\sim 140$ MeV) dominated by the leptons. The rounded down temperature of $100~\text{MeV}$ is a good definition for the start of the lepton era. From here on the universe is mainly filled with electrons, positrons, the 6 neutrinos, and the photons. At around $1~\text{MeV}$ ($10^{10}~\text{K}$), two coinciding events take place: the neutrinos decouple from matter, and soon after the temperature drops below the threshold for e$^-\text{e}^+$ production, so they will start to annihilate. As the last positrons die out we enter the plasma era, where the universe is filled with a hot plasma of photons and the residual matter consisting of nuclei and electrons.

\subsection{Constraints in our model} 

For our model we use theories which is constrained to the lepton era. Hence its validity drops as we cross over from both sides: going to higher temperatures leading into the hadron era, and dropping to lower temperatures leading into the plasma era. Expanding out of the (more) valid lepton-era can still show us some interesting physics worth exploring. We use the following constraints/simplifications in our model:

\begin{description}[leftmargin=*]

\item[A hadron-less universe:] At temperatures of 100 MeV ($10^{12}$ K) the hadronic content of the universe is mainly pions ($m_{\pi^0} = 135.0 $ MeV/c$^2$, $m_{\pi^\pm} = 139.6 $ MeV/c$^2$) and only to a smaller extent other heavier hadrons, like kaons, protons and neutrons. Going above the phase transition at $T \approx 170$ MeV there will be a quark-gluon plasma with all its implications. This requires taking QCD theory into account. For simplicity we disregard all hadrons. This will give a better understanding of the consequences by adding a single particle pair to our model.

\item[Matter-antimatter neutral universe:] Data from WMAP \citep{WMAP:2012} suggest a baryon-to-photon ratio  of $(6.079 \pm 0.090)\times 10^{-10}$, which roughly corresponds to the matter over antimatter difference. This hardly plays any role when the universe is still hot enough to produce particle pairs, but this asymmetry have to be taken into consideration as we enter the plasma era, when all anti-matter is annihilated, and we are left with the residual matter.

\item[No neutrino decoupling:] Neutrinos have a very long mean free path, which makes them the primary source for momentum transfer. Being the main contributor to both shear and bulk viscosity, an accurate model of the neutrino density is important. As the rate of the (weak) neutrino interactions becomes slower than the rate of expansion of the universe the neutrinos will decouple. This happens at $\Gamma_{\nu}/H \approx [\GF^2/(\hbar c)^6] \sqrt{\hbar c / \GN}(\kB T)^3 \simeq (T / 10^{10} \text{ K})^3$ (see Sect. \ref{sec:NeutrinoDecoupling}). We have neglected the neutrinos in our calculations, but marked the decoupling temperature in our figures.

\item[No electromagnetic contribution:] Particle interaction in the lepton era is dominated by the weak and electromagnetic forces. Viscous effects are related to momentum transfer. The long mean free path of the neutrinos ensures that its momentum transfer is several orders of magnitude larger to those of electromagnetic origin during this era. Electromagnetic interactions do however play a role at low temperatures after the neutrinos decouples (plasma era) and at high temperatures when the weak and electromagnetic forces are closer in strength (electroweak scale). This argument should be valid at temperatures  beneath $10^{13}$ K, but become more questionable as we get towards $10^{14}$ K \citep{Hoogeveen:Epsilon}.

\item[No reheating from annihilations:] 
The decay of heavier particles to lighter particles result in a temperature increase, as the rest masses are converted to kinetic energy. Any out-of-equilibrium effects of this origin has not been considered in this paper. 

\item[Elastic collisions and no chemical potential:]
As is done in earlier writings, we only use elastic collisions in our calculations. The cross-section of inelastic collisions is of comparable sizes, so this approximation should not affect the results considerably \citep{Hoogeveen:1986}. The chemical potentials of all particles is set to zero.
\end{description}

\section{Viscous theory}
\label{Sec:ViscousTheory}
Two factors which play a role in viscosity are, first the state of the system --- which for our case is the particle composition of the system, and secondly the transport equation for the system. We use the theory as described by \cite{Hoogeveen:1986}. The core theory and equations for shear ($\eta_s$) and bulk ($\eta_v$) viscosities are given below, and a more thorough description of the linear equations is described in Appendix \ref{App:Coefficients}. First of, the shear and bulk viscosities are found using the Chapman-Enskog approximation \citep{vanErkelens:1978, deGroot:RelativisticKineticTheory}:
\begin{align}
  \eta_s &= \frac{1}{10} cn \left( \frac{\kB T}{c} \right)^3 \sum_k c_k \gamma_k,
  \label{Eq:EtaS}\\  
  \eta_v &= n\kB T \sum_k a_k \alpha_k,
  \label{Eq:EtaV}
\end{align}
where $c$ is the speed of light, $n$ is total particle density, $\kB$ is Boltzmann's constant, $T$ is temperature. The coefficients $c_k$ and $a_k$ originate from the linearized relativistic Boltzmann equations. These transport equations describe the energy-momentum transfer, essentially the ``force" trying to drive the system back to equilibrium. The two sets of linear equations are given in Appendix \ref{App:Coefficients} for particle $k$. $\gamma_k(n_k,T)$ and $\alpha_k(n_k,T)$ are functions of the particle densities and temperature and describe the state of the system. For the shear viscosity case this is a product of the particle composition (fraction) and enthalpy of the system, while for the bulk case it is a more complex function of these. The two
functions are given by:
\begin{equation}
  \gamma_k = 10 x_k \hat{h}_k,
  \label{Eq:Gamma}
\end{equation}
\begin{equation}
  \alpha_k = -x_k \left[ \gamma - (\gamma - 1)(z_k^2 + 5\hat{h}_k -\hat{h}_k^2) \right],
  \label{Eq:Alpha}
\end{equation}
where $x_k$, $z_k$ and $\hat{h}_k$ are particle fraction, inverse dimensionless temperature and dimensionless partial specific enthalpy respectively, and are defined as:
\begin{subequations}
  \begin{equation}
    x_k \equiv \frac{n_k}{n},
    \label{Eq:x}
  \end{equation}
  \begin{equation}
    z_k \equiv \frac{m_k c^2}{\kB T},
    \label{Eq:z}
  \end{equation}
  \begin{equation}
    \hat{h}_k \equiv z_k \frac{K_3(z_k)}{K_2(z_k)}.
    \label{Eq:hHat}
  \end{equation}
\end{subequations}
Here $K_3$ and $K_2$ are the modified Bessel functions of the second kind. $(z_k^2 + 5\hat{h}_k -\hat{h}_k^2)$ is actually the specific heat capacity per particle for constant pressure, $\hat{c}_P$. The $\gamma$ in Eq. (\ref{Eq:Alpha}) is the heat capacity ratio (also known as the adiabatic index) and is given by\footnote{From \cite{deGroot:RelativisticKineticTheory} we have $\gamma = c_P/c_V = \left( \frac{\partial h}{\partial T} \right)_P / \left( \frac{\partial \ek}{\partial T} \right)_V$, where $\ek=mc^2\frac{K_1(z)}{K_2(z)}+3\kB T = mc^2\frac{K_3(z)}{K_2(z)}-\kB T$ and $h=\ek+\frac{P}{n} = mc^2\frac{K_3(z)}{K_2(z)}$.} 
\begin{equation}
  \frac{\gamma}{\gamma -1} = \sum_{k=1}^N x_k (z_k^2 + 5\hat{h}_k -\hat{h}_k^2).
  \label{Eq:GammaInAlpha}
\end{equation}

For low temperatures the heat capacity ratio should coincide with the classical value for a non-relativistic gas and approach 5/3. For massless particles, the average kinetic energy per particle is $3\kB T$, the enthalpy per particle is $4\kB T$, such that $\gamma$ should approach $4/3$ in the relativistic limit. We look closer at $\hat{c}_P =(z^2 + 5\hat{h} -\hat{h}^2)$, $\alpha_k$, and their properties in Fig. \ref{Fig:AlphaDetailed}. The particle fractions $x_k$, depends solely on
the particle densities.

\subsection{Particle densities}
The particle density for a particle $k$ is given by \cite{Andersen:StatMech, Hoogeveen:1986} as
\begin{subequations}
\begin{equation}
  n_k(T)=\frac{g_k}{2 \pi^2 \hbar^3} 
  \int_{m_kc^2}^\infty \frac{E \sqrt{E^2-m_k^2c^4}}{e^{E/\kB T}-\delta_k}dE,
  \label{Eq:nk}
\end{equation}
where $g$ is the number of spin states and $\delta$ is $+1$ for bosons and $-1$ for fermions, which gives us the expressions for the 13 different particles:
\begin{align}
  n_{e^\pm \mu^\pm \tau^\pm}(T) &= \frac{1}{\pi^2} \left( \frac{\kB T}{\hbar c} \right)^3 \int_z^\infty \frac{u\sqrt{u^2-z^2}}{e^u+1}du, \label{Eq:nl}\\
  n_{\gamma}(T)     &= \frac{1}{\pi^2} \left( \frac{\kB T}{\hbar c} \right)^3 2\zeta(3), \label{Eq:ng}\\
  n_{\nu_e \bar{\nu}_e \nu_\mu \bar{\nu}_\mu \nu_\tau \bar{\nu}_\tau}(T) 
         &= \frac{1}{\pi^2} \left( \frac{\kB T}{\hbar c} \right)^3  \frac{3}{4} \zeta(3) \label{Eq:nv},
\end{align}
\end{subequations}
where $\zeta(3) \approx 1.202$ is the Riemann zeta function and $u$ (just as $z$) is a dimensionless inverse temperatures defined by
\begin{equation}
  u \equiv \frac{E}{\kB T}.
  \label{Eq:u}
\end{equation}

\begin{figure}[h]
  \centering
  \includegraphics[width=\columnwidth]{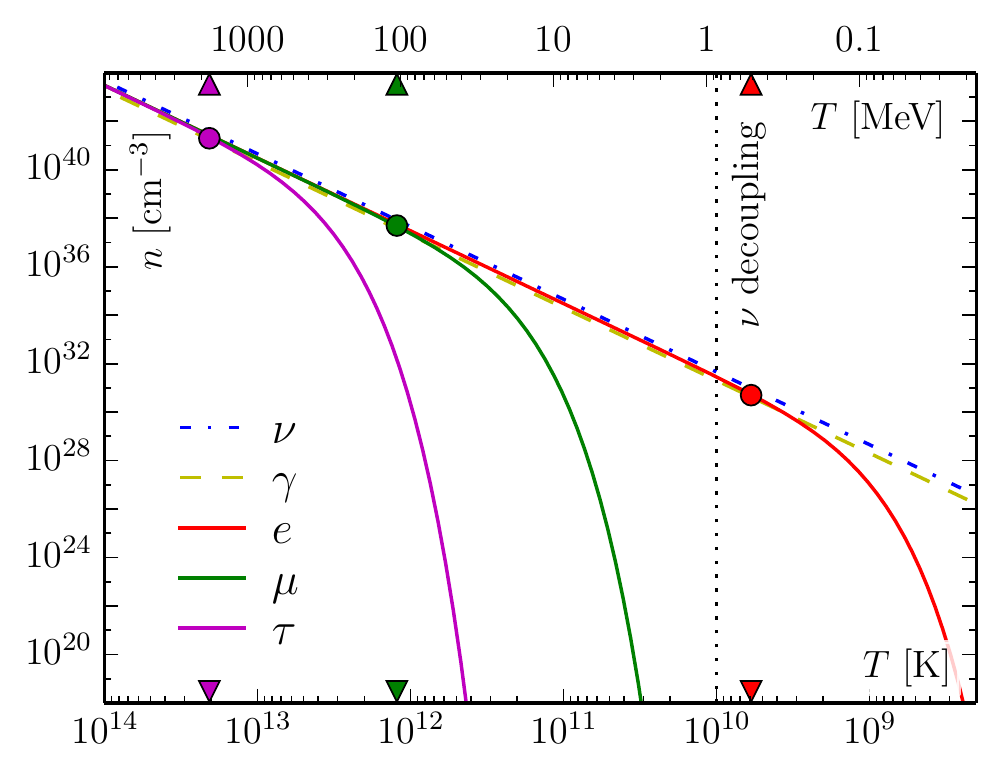}
  \caption{Particle densities $n_k$ as functions of temperature. Similar particles (i.e. 1 photon, 6 neutrinos, 2 electrons, 2 muons and 2 taus) are added together. Equivalent rest mass of particles are marked by triangles and circles.
  }
  \label{Fig:ParticleDensities}
\end{figure}

\begin{figure}[h]
  \centering
  \includegraphics[width=\columnwidth]{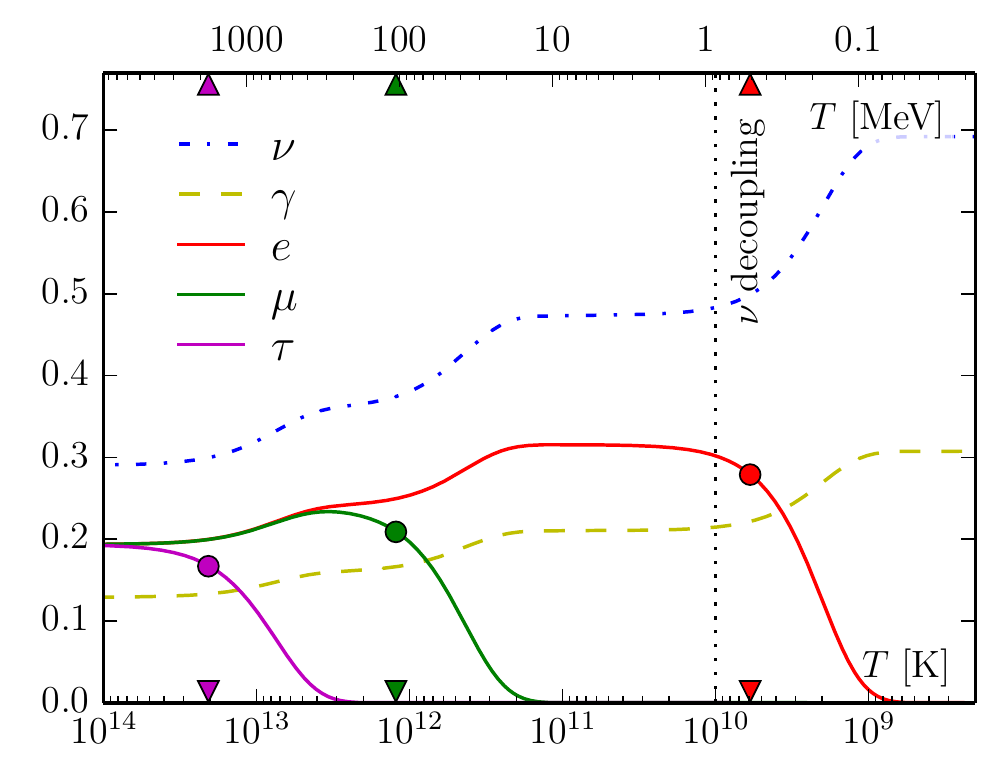}
  \caption{Particle fractions $x_k$ as functions of temperature.}
  \label{Fig:ParticleFractions}
\end{figure}

Using dimensionless numbers in Eq. (\ref{Eq:nk}), the integrals themselves will vary from 0 (low temperature) to $3\zeta(3)/2$ for fermions and $2\zeta(3)$ for bosons at high temperatures. The numerical solutions for the particles densities are given in Fig. \ref{Fig:ParticleDensities}, and that for the particle fractions in Fig. \ref{Fig:ParticleFractions}. It is worth noting how late the pair production of massive particles is maintained. Their contribution is significant at temperatures well below the pair production energy threshold of 2$m_k$. 

\subsection{Bulk viscosity and its terms}

\begin{figure*}
 \centering
	\includegraphics[width=\textwidth]{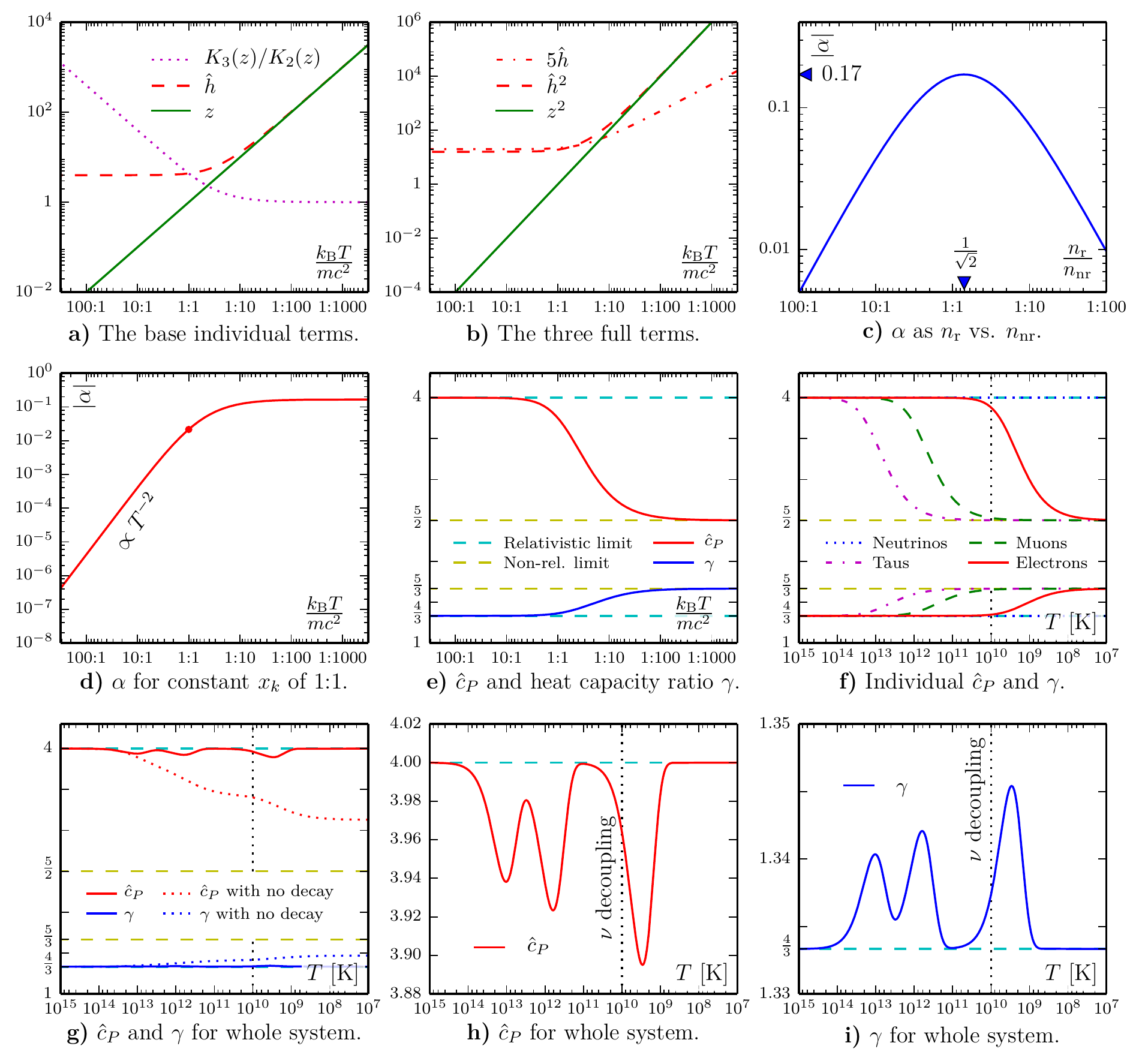}
 \caption{A more thorough look at the specific heat capacity per particle ($\hat{c}_P$), the heat capacity ratio ($\gamma$), and $\alpha$, as they are decomposed into its basic constituents. 
  Subfigures \textbf{(a)} and \textbf{(b)} shows functions of $z$, $\hat{h}$ and the Bessel functions, which make up $\hat{c}_P$. 
  \textbf{(c)} shows $\alpha$ as a function of the amount of relativistic to non-relativistic particles. The momentum transfer is biggest when ratio is $2^{-1/2}$. 
  \textbf{(d)} shows a system with one massless and one massive particle specie in equal amount. $\alpha$ will increase as the massive particles go from being relativistic to non-relativistic.
  \textbf{(e)} shows $\hat{c}_P$ and $\gamma$ as function of $k_\textbf{B}T/mc^2$, and
  \textbf{(f)} for the three charged leptons as function of temperature.
The last three subfigures shows the complete functions of $\hat{c}_P$ and $\gamma$. 
  For \textbf{(g)} we have included a hypothetical scenario with no particle annihilation (decay). $\hat{c}_P$, shown in \textbf{(h)} drops almost 7\% towards the non-relativistic limit of $5/2$, while $\gamma$, shown in \textbf{(i)} rise almost 3.5\% towards the non-relativistic of $5/3$.}
 \label{Fig:AlphaDetailed}
\end{figure*}

Let us revisit the concept of bulk viscosity by giving a few examples. During expansion relativistic and a non-relativistic gases drop differently in temperature. For relativistic gases we have $T \propto a^{-1}$, while the case for non-relativistic gases is $T \propto a^{-2}$. Since non-relativistic gases drop quicker in temperature we will have heat transfer from the warmer relativistic gas to the colder non-relativistic gas in order to regain a thermal equilibrium in the mixture. Bulk viscosity increases with the temperature difference. Neutrinos have a long mean free path, hence longer mean life time than the charged leptons and the photon which interact much stronger, and hence the neutrinos are the primary contributor to the bulk viscous term.

As mentioned earlier, the $\alpha_k$ coefficients tells us about the state of the system, more specifically it tells us about the ratio of particles which wants to drop faster (non-relativistic particles) in temperature to those who wants to drop slower (relativistic particles). Particles, or the particle specie as a whole to be more specific, with a positive $\alpha_k$ are above the equilibrium temperature, and the other way around for particles with a negative $\alpha_k$, and the sum of all $\alpha_k$'s is zero. Since massive particles gradually change from behaving
(ultra-)relativistic, to non-relativistic, and finally annihilation, the ``sweet spot" should therefor be somewhere in between. Looking at Fig. \ref{Fig:AlphaDetailed}c) we see this peak in $\alpha$ when the ratio of non-relativistic (nr) to relativistic particles (r) is $\sqrt{2}:1$. For $n_r \gg n_{nr}$ we have $\alpha \propto n_r / n_{nr}$, and will decrease in the same manner for $n_r \ll n_{nr}$. For ratios of $1:1$ and $2:1$ $\alpha$ will be $1/6$, while the peak at $\sqrt{2}:1$ will be just slightly larger.

\begin{figure*}
  \centering
  \includegraphics[width=\textwidth]{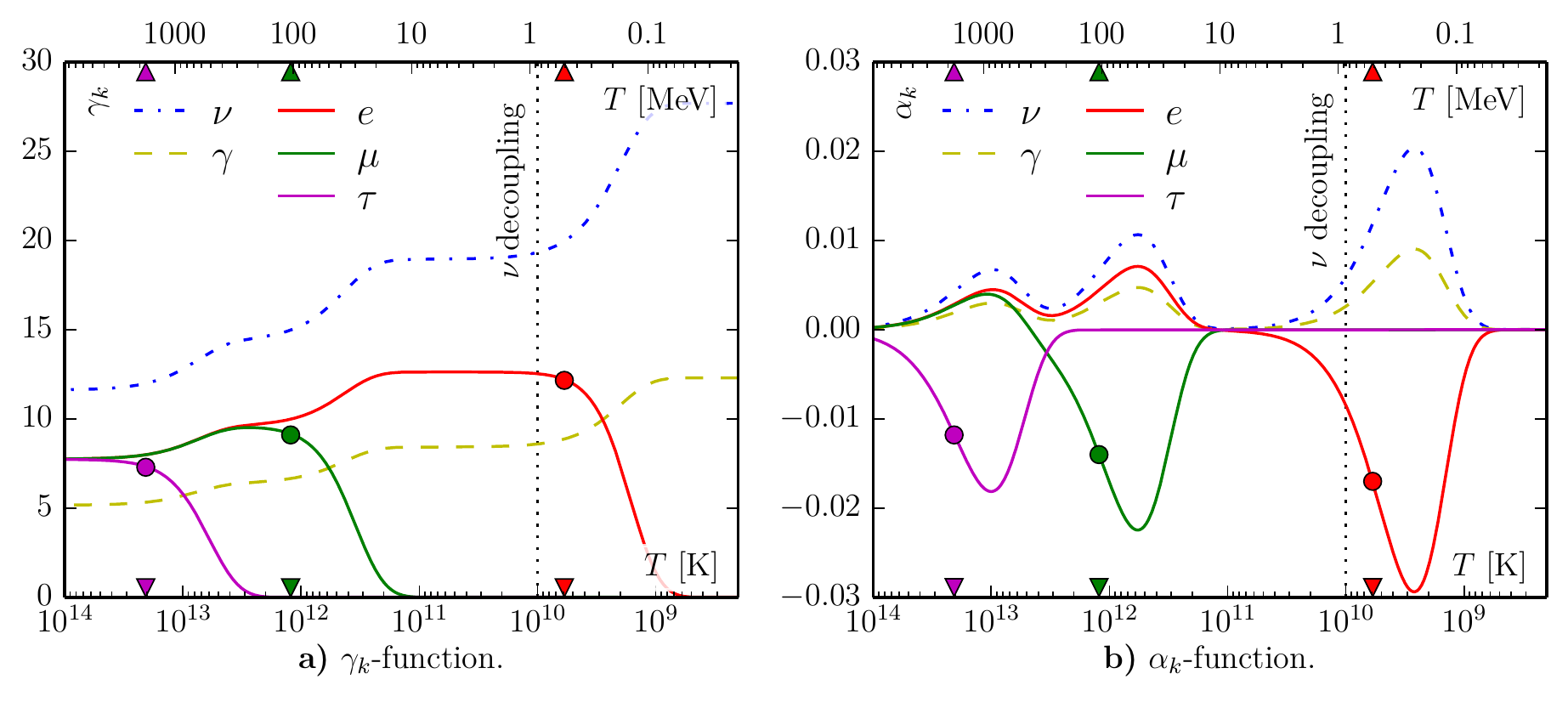}
  \caption{$\gamma_k$ \textbf{(a)} and $\alpha_k$ \textbf{(b)} as functions of temperature. We clearly see how $\gamma_k$ correlates to the particle fraction plot of Fig. \ref{Fig:ParticleFractions} (just slightly shifted to the left). $\alpha_k$ peaks at temperatures where there is a good portion of both relativistic particles and non-relativistic particles. In practice this is the periods from when, in order, the tau, muon and electron particles become non-relativistic until they annihilate.}
  \label{Fig:GammaAndAlpha}
\end{figure*}

In Fig. \ref{Fig:AlphaDetailed}d) we have shown a case with one massive and one massless particle with no particle decay (such that each particle number stays constant). At high temperature most of the massive particles move at relativistic velocities, and the mixture expands with negligible bulk viscosity. As the temperature drop, a bigger and bigger portion of the massive particles will behave more non-relativistic, and $\alpha$ will grow as $T^{-2}$ until we reach a temperature roughly equal that of the particle mass. The growth in $\alpha$ will slow down and converge towards 1/6 as the ratio of relativistic to non-relativistic particles goes towards $1:1$. The complete functions of $\gamma_k$ and $\alpha_k$ are plotted in Fig. \ref{Fig:GammaAndAlpha}.

\subsection{Viscous pressure} 

For a system in thermodynamic equilibrium, the energy density\footnote{As not to be confused with the mass density $\rho = \epsilon /c^2$} $\epsilon$ and pressure $P$ is found by adding together the individual pressures from all particle types $k$:
\begin{equation}
  \epsilon(T) = \sum_k \frac{g_k}{2 \pi^2 \hbar^3} \int_{m_kc^2}^\infty \frac{E^2 \sqrt{E^2-m_k^2c^4}}{e^{E/\kB T} \pm 1}dE ,
  \label{Eq:EnergyDensity}
\end{equation}
\begin{equation}
  P(T) = \sum_k \frac{g_k}{6 \pi^2 \hbar^3} \int_{m_kc^2}^\infty \frac{(E^2-m_k^2c^4)^{3/2}}{e^{E/\kB T} \pm 1}dE,
  \label{Eq:Pressure}
\end{equation}
Eq. \ref{Eq:Pressure} is normally written in a simplified version as
\begin{equation}
  P(T) = g_{\star_P}(T) \frac{\pi^2}{90}\frac{(\kB T)^4}{(\hbar c)^3} \equiv P_\text{TE},
  \label{Eq:PressureThermodynamic}
\end{equation}
where $g_{\star_P}(T)$ is the relativistic degrees of freedom for pressure (compared to that of the photon). For periods where the particle fractions stay constant the degrees of freedom stay constant as well (for example just before neutrino decoupling $g_{\star_P} \simeq 10.75$). The evolution of $g_{\star_P}$ for all particles in the standard model is plotted in the upper right corner of Fig. \ref{Fig:GstarPressure}. We will define this thermodynamic equilibrium pressure for $ P_\text{TE}$, making the temperature dependence implicit.

The viscous pressure is \citep{Brevik:2005vc}:
\begin{equation}
P_\text{visc} = -\eta_v u^\mu_{;\mu} = -3\eta_v H,
\end{equation}
where $H$ is the Hubble constant given by 
\begin{equation}
H = \sqrt{\frac{8\pi\GN}{3} \rho} = \sqrt{\frac{8\pi\GN}{3} \frac{\epsilon}{c^2}}.
\end{equation}
The effective pressure is found by
\begin{equation}
P_\text{eff} = P_\text{TE} + P_\text{visc} = \frac{\epsilon}{3} - 3\eta_vH.
\end{equation}

\subsection{Neutrino decoupling}
\label{sec:NeutrinoDecoupling}

As the rate of the neutrino interactions, $\Gamma_{\nu}$, becomes slower than the rate of expansion of the universe, $H$, the neutrinos will decouple. The collision rate of neutrinos with electrons and positrons is \citep{Weinberg:Cosmology}:
\begin{equation}
\begin{aligned}
    \Gamma_{\nu}	&= n_\text{e} \sigma_{\text{wk}} 
    \approx \left( \frac{\kB T}{\hbar c} \right)^3 (\hbar c \Gw \kB T)^2 \\
    &\approx \frac{\Gw^2(\kB T)^5}{\hbar c},
\end{aligned}    
\label{Eq:NeutrinoElectronCollisionRate}
\end{equation}
where $n_e$ is the number density of electrons and $\sigma_\text{wk}$ is the neutrino-electron scattering. $\Gw = \GF/(\hbar c)^3 \approx 1.166 \times 10^{-5}$ GeV$^{-2}$ \citep{Griffiths:Particles, PDG:2012} is the weak coupling constant.
By using the energy density as given in Eq. (\ref{Eq:EnergyDensity}), the expansion rate at the same time is
\begin{equation}
\begin{aligned}
H &= \sqrt{\frac{8\pi \GN}{3c^2} \epsilon} 
  = \sqrt{\frac{8\pi \GN}{3c^2} g_\star \frac{\pi^2}{30}\frac{(\kB T)^4}{(\hbar c)^3}} \\
  &\approx \sqrt{\frac{\GN(\kB T)^4}{(\hbar c)^3}}.
\end{aligned}  
\end{equation}
where $g_\star$ is the effective degrees for freedom for energy density.
The prefactors in $\Gamma_\nu$ and $H$ roughly cancel each other out, such that we end up with
\begin{equation}
  \frac{\Gamma_{\nu}}{H} \approx \Gw^2 \sqrt{\frac{\hbar c}{\GN}}(\kB T)^3 
  \simeq \left(\frac{T}{10^{10} \text{ K}} \right)^3.
  \label{Eq:NeutrinoDecouplingTemperature}
\end{equation}
It is worth noting that without taking neutrino decoupling into account, our equations would result in the shear viscosity going towards infinity as $T$ goes towards zero. However the bulk viscosity would still drop as it relies on the ratio of relativistic to non-relativistic particles.

\section{Numerical model}
\label{Sec:NumericalModel}
Our model is made with \textit{Mathematica 9}, using standard precision. We calculate the shear viscosity for 6, 9, 11 and 13 particles. For bulk viscosity we use 9, 11 and 13 particles. This refers to: pure neutrino model ($\nu_e,\bar{\nu_e},\nu_\mu,\bar{\nu_\mu},\nu_\tau,\bar{\nu_\tau} = 6$), plus electron/positron and photon (plus $e^-,e^+,\gamma = 9$), plus muons (plus $\mu^-,\mu^+ = 11$), plus taus (plus $\tau^-,\tau^+ = 13$). We use the particle and physical constants from PDG's ``Review of Particle Physics \citep{PDG:2012}. The Weinberg angle (\W) is set to 0.229, as it was in the original 1986 paper. The expanded parameters including $\tau^-, \tau^+$ for differential cross-sections in elastic collisions are given in Appendix \ref{App:CrossSections}, Table \ref{Tab:va}. 

Using these new parameters, we have solved the coefficients as given in Appendix A in \citep{Hoogeveen:1986}. The equations used are reprinted in Appendix \ref{App:Coefficients}.

Some integrals have been solved numerically instead of analytically. This has resulted in some high precision errors (e.g. when $\alpha_k$ goes towards zero). This affected the calculations at low temperatures when including the tau and muon particles. The results did however converge to the ``2-particle-less" models beforehand, and hence these errors can be disregarded. For low temperature  where these are the cases, the cells in Table \ref{Tab:ViscosityTableSmall} are left blank.

The particle densities, and the $\alpha_k$ and $\gamma_k$ terms were calculated and plotted continuously, the viscosity terms themselves consist of a series of single calculations over 74 different temperatures and 7 particle models (4 for shear and 3 for bulk viscosities).

Readers interested in the \textit{Mathematica} code are encouraged to email the author.

\section{Results}
\label{Sec:Results}

\begin{figure}[h]
	\centering
	\includegraphics[width=\columnwidth]{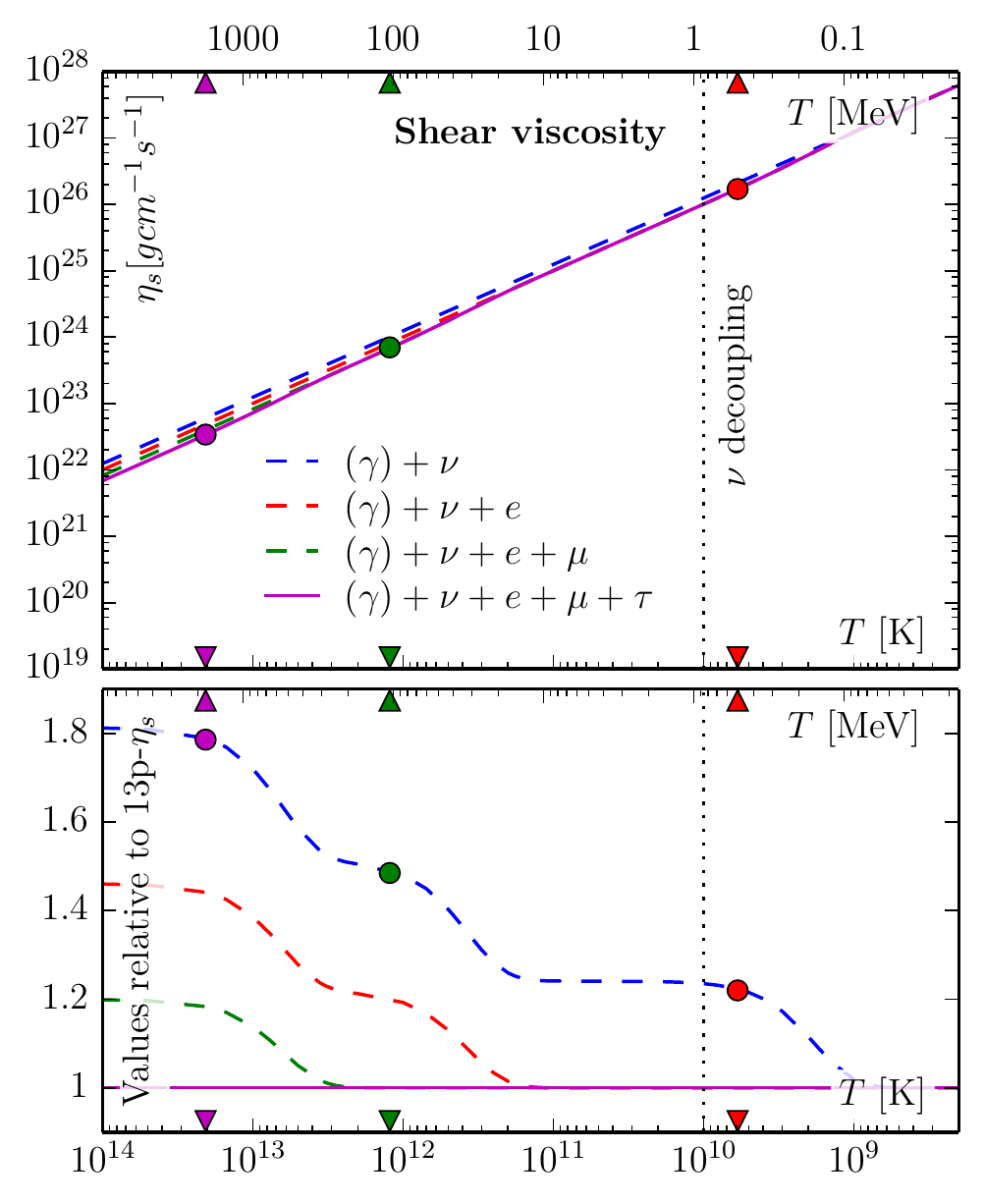}
	\caption{\textit{Top}: Absolute value of $\eta_s$. Its value increases slightly more rapidly in the regions where heavier particles annihilate. In ``particle-stable" regions $\eta_s$ goes as $\propto T^{-1}$. The \textit{bottom plot} shows the deviations from  the 13 particle model.}
	\label{Fig:EtaS}
\end{figure}

\begin{figure}[h]
	\centering
	\includegraphics[width=\columnwidth]{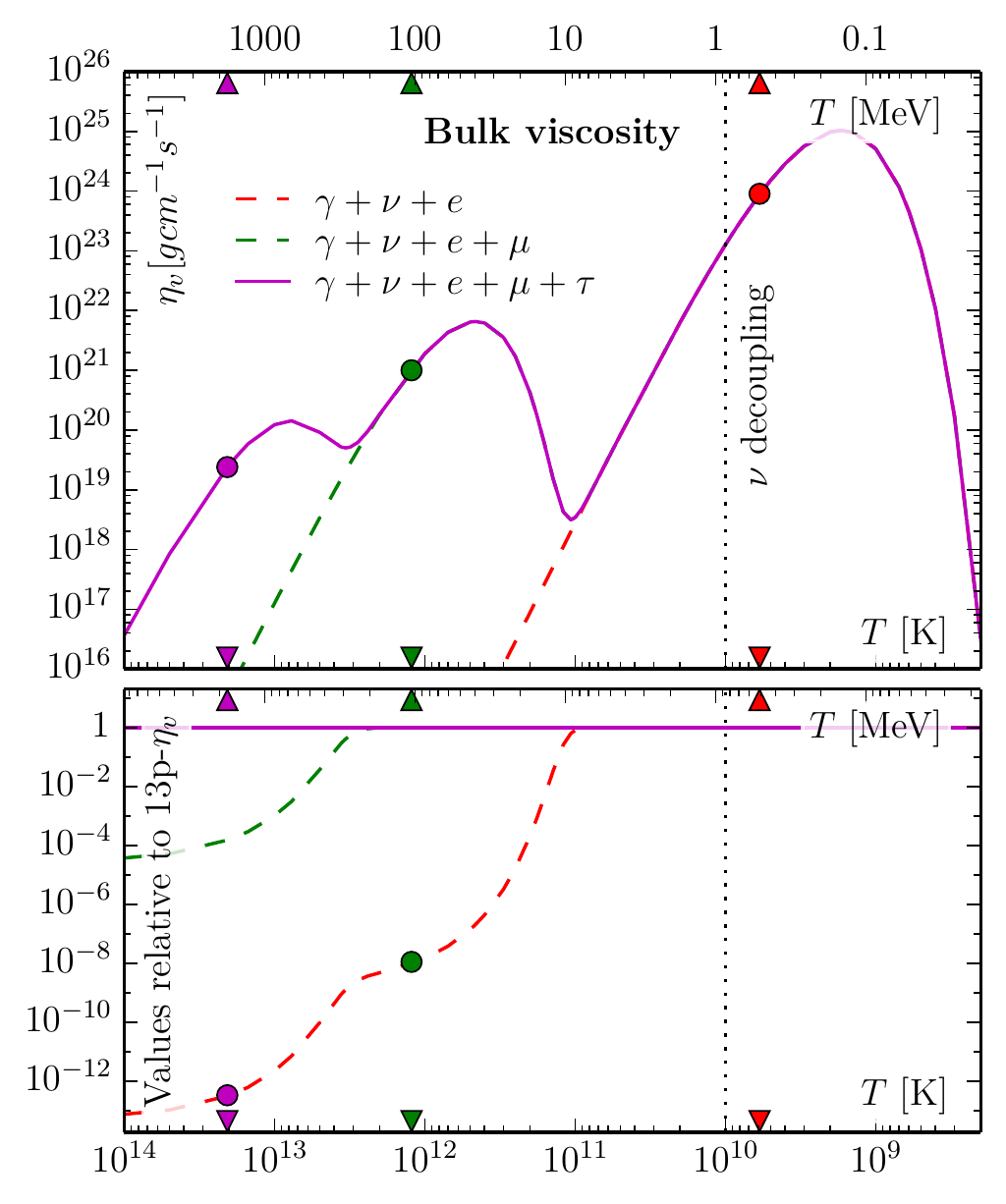}
	\caption{\textit{Top plot} show the bulk viscosity, $\eta_v$. In the particle-fraction-stable regions the bulk viscosity increases rapidly as $\eta_v \propto T^{-5}$, and then drop quickly as the heavier particles dies out. The local maxima's (peaks) corresponding to the three lepton generations show that three peaks in $\eta_v$ goes as $\propto T^{-4/3}$. The \textit{bottom plot} shows the deviations from  the 13 particle model.}
	\label{Fig:EtaV}
\end{figure}

All calculations in this article are given in CGS units. Our main results are given in Table \ref{Tab:ViscosityTableSmall}, where we have organized the shear and bulk viscosities by numbers of particles used in the model. For shear viscosity, this is with 7, 9, 11 and 13 particles. Since photons have no impact on the shear viscosity, we might as well have written 6, 8, 10 and 12. As bulk viscosity requires both a relativistic and a non-relativistic part, we only have 9, 11 and 13 particle models. Our results are graphed in Fig. \ref{Fig:EtaS}, \ref{Fig:EtaV}, ranging from $2\times10^8$ to $10^{14}$ K ($\approx 20$ keV to 10 GeV). Everything below the neutrino decoupling temperature should be considered non-physical. We have included them in our calculations and plot as they still give us some input on the behavior of the the two viscous terms.

For Table \ref{Tab:ViscosityTableSmall} the range is increased one order of magnitude to $10^{15}$ K (which is uncomfortable close to the electroweak scale).

The peaks in $\alpha_k$ and related peaks in the bulk viscosity, $\eta_v$, are given in Table \ref{Tab:Peaks}. The associated electron peaks are found at 0.44 $m_e$ for $\alpha$ and 0.29 $m_e$ for $\eta_v$. For the muon and tau particles they both peak at 0.47 $m_{\mu,\tau}$ for $\alpha$ and 0.37 $m_{\mu,\tau}$ for $\eta_v$. 

\begin{figure}
	\centering
	\includegraphics[width=\columnwidth]{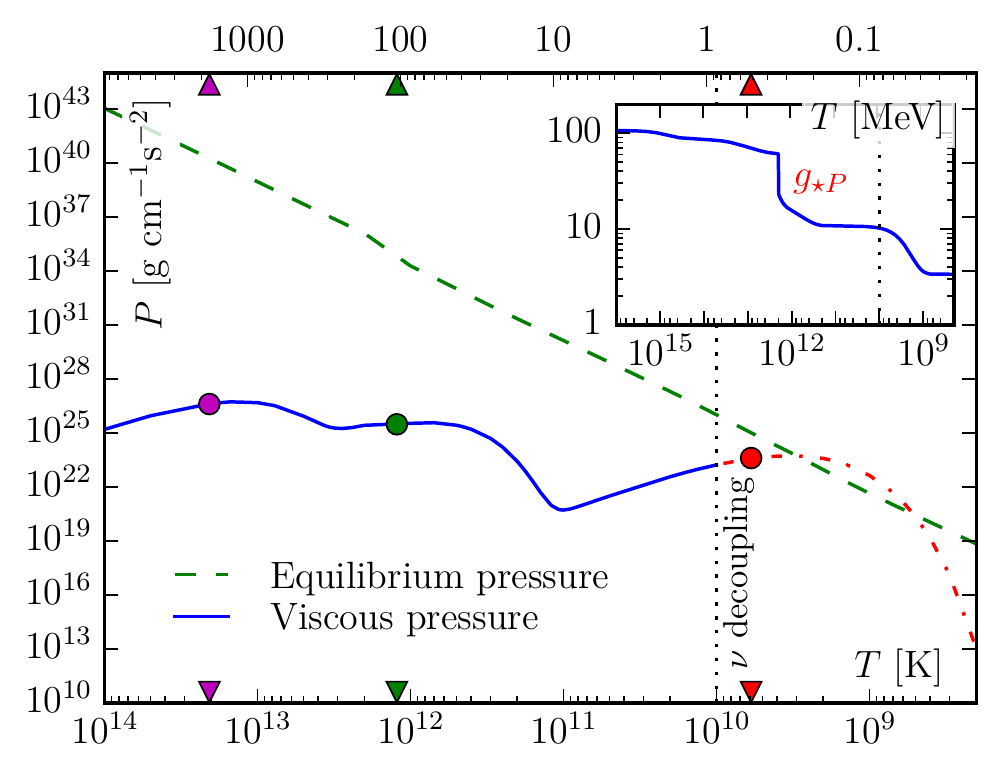}
	\caption{Thermodynamic equilibrium pressure and viscous pressure.
The calculations for the latter pressure does not make sense after the
neutrinos decouple and are plotted as \textit{red dash-dots}. The thermodynamic pressure $P_\text{TE}$ is a function of $g_{\star P}$ (shown in the subplot) and $T$.}
	\label{Fig:GstarPressure}
\end{figure}

Our results using a 9-particle model are very similar to those found by Hoogeveen et al., only differing by around 1\% - which is within uncertainty. However, for our more particle-rich models the differences are quite significant in the high temperature regions. Plotting the pressures (Fig. \ref{Fig:GstarPressure}) shows that the viscous pressure has a negligible contribution at high temperatures, and is at its highest at the time of the neutrino decoupling (and is actually larger than the the thermodynamic pressure if we  disregard the decoupling).

The rest of our results from Fig. \ref{Fig:EtaS}, \ref{Fig:EtaV} and Table \ref{Tab:ViscosityTableSmall} will be divided into a shear and bulk viscosity part. As a side note we should say that by using PDG's \citep{PDG:2012} entry for \W will keep $\alpha_k$ unchanged, but the values of $\eta_{s}$ and $\eta_{v}$ will decrease by roughly 1\%.

\subsection{Shear viscosity}
\label{Sec:EtaS}

\begin{itemize}[noitemsep]
\item The photon contribution to shear viscosity is negligible, so removing them and making models for 6, 8, 10 and 12 particles gives the same results.
\item Shear viscosity goes as $T^{-1}$ as long as the particle fractions stays constant. This can  be understood by the fact that energy density (for relativistic particles) goes as $T^4$ while the interaction rate, $\Gamma$ goes as $T^5$. As shear viscosity depends on the mean free path of the particles, this is the inverse of the interaction rate.
\item At high temperatures where all charged leptons are present we get a drop of $\eta_s$ by including the extra leptons in our model. For every particle pair we include in our model, $\eta_s$ drops by rougly 20\%, or more specifically
\begin{equation}
\eta_{s(7p)} \xrightarrow{19\%} \eta_{s(9p)} \xrightarrow{18\%} \eta_{s(11p)} \xrightarrow{17\%} \eta_{s(13p)}.
\end{equation}
\end{itemize}

\subsection{Bulk viscosity}
\label{Sec:EtaV}

\begin{itemize}[noitemsep]
\item Bulk viscosity is much more sensitive to temperature and number densities of the different particles.
\item Photons contribute to bulk viscosity indirectly as they heat up the non-relativistic leptons. Removing the photon from our model would decreased $\eta_v$ by 38\%, 29\% and 24\% for our 9, 11 and 13 particle models.
\item Bulk viscosity goes as $T^{-5}$ as long as the particle number stays constant. When a particle species dies out $\eta_v$ drops as the particle density function (which is exponential decay).
\item At high temperatures where all charged leptons are present we get a significant increase in $\eta_v$ by including the extra leptons in our model. Specifically the two jumps are:
\begin{equation}
\eta_{v(9p)} \xrightarrow{4.88\E9} \eta_{v(11p)} \xrightarrow{3.22\E4} \eta_{v(13p)}.
\end{equation}
\item The peaks for the local maximas of $\eta_v$ goes roughly as $T^{-4/3}$ (within 6\%). The maximas and minimas are marked by $\dagger$ and $\ddagger$ in Table \ref{Tab:ViscosityTableSmall}. (A newer value for \W from PDG \citep{PDG:2012} will shift the local maxima and minimum peaks by less than 0.1\%).
\end{itemize}
\begin{itemize}[noitemsep]
\item To understand why the peaks in bulk viscosity are as they are we have to look at several factors. Just as shear viscosity, the interacting rate vs. energy density should go as $T^{-1}$. To give an example of this: If the electron was 10 times heavier, the bulk viscosity would be a 10 times lower. That is: Changing the masses of particles changes the peaks as $T^{-1}$.
\item The remaining $T^{-4}$ comes from the product of the transport equations ($a_k$) and the $\alpha_k$ parameters. Each of which goes as $T^{-2}$.
\end{itemize}

\begin{table}[h]
\centering
\begin{small}
\subtable[$\alpha_k$ related peaks]
 {\begin{tabular}{ll@{\,\,}rl@{\,\,}rl}
 \tableline
 $e$ peak 		& $T$ = 2.642\E9  & K & 227.7 & keV & 0.4455 $m_e$ \\
 $\mu$ peak		& $T$ = 5.763\E11 & K & 49.66 & MeV & 0.4700 $m_{\mu}$\\
 $\tau$ peak	& $T$ = 9.718\E12 & K & 837.4 & MeV & 0.4713 $m_{\tau}$\\
 \end{tabular}}
\subtable[$\eta_v$ related peaks]
 {\begin{tabular}{ll@{\,\,}rl@{\,\,}rl}
 $e$ peak 		& $T$ = 1.698\E9  & K & 146.3 & keV	& 0.2863 $m_e$\\
 $\mu$ peak		& $T$ = 4.597\E11 & K & 39.61 & MeV & 0.3749 $m_{\mu}$\\
 $\tau$ peak	& $T$ = 7.710\E12 & K & 664.4 & MeV & 0.3739 $m_{\tau}$\\
 \tableline 
 \end{tabular}}
\end{small}
\caption{Local maxima peaks in $\alpha_k$ and $\eta_v$.}
\label{Tab:Peaks}
\end{table}

\begin{table*}[h]
\begin{small}
\begin{center}
\begin{tabu}{ll|cccc|lcc}
&& \multicolumn{4}{c|}{$\eta_s$ [g/cm\,s]} & \multicolumn{3}{c}{$\eta_v$ [g/cm\,s]}\\
\multicolumn{1}{c}{$T$ (K)} 	& \multicolumn{1}{c|}{$T$ (eV)}	 & 
\multicolumn{1}{c}{7p}	& \multicolumn{1}{c}{9p}	& \multicolumn{1}{c}{11p} 	& \multicolumn{1}{c|}{13p} & 
\multicolumn{1}{c}{9p}	& \multicolumn{1}{c}{11p} 	& \multicolumn{1}{c}{13p} 								\\
	\hline
\rowfont{\color{orange}}
1.000\E15 & 86.17 GeV & 1.241\E21 & 9.997\E20 & 8.208\E20 & 6.843\E20 & 2.824\E-2 & 1.379\E7\phantom{0} & 4.445\E11 \\
\rowfont{\color{orange}}
1.000\E14 & 8.617 GeV & 1.241\E22 & 9.997\E21 & 8.208\E21 & 6.846\E21 & 2.823\E3  & 1.376\E12 & 3.658\E16 \\
5.000\E13 & 4.309 GeV & 2.481\E22 & 1.999\E22 & 1.642\E22 & 1.371\E22 & 9.033\E4  & 4.385\E13 & 8.423\E17 \\
2.000\E13 & 1.723 GeV & 6.204\E22 & 4.998\E22 & 4.104\E22 & 3.469\E22 & 8.821\E6  & 4.182\E15 & 2.656\E19 \\
1.000\E13 & 861.7 MeV & 1.241\E23 & 9.997\E22 & 8.209\E22 & 7.213\E22 & 2.823\E8  & 1.260\E17 & 1.227\E20 \\
7.710\E12{$^\dagger$\hspace{-10pt}} 
		  & 664.4 MeV & 1.609\E23 & 1.297\E23 & 1.065\E23 & 9.613\E22 & 1.036\E9  & 4.423\E17 & 1.428\E20 \\
5.000\E12 & 430.9 MeV & 2.481\E23 & 1.999\E23 & 1.643\E23 & 1.564\E23 & 9.033\E9  & 3.419\E18 & 9.135\E19 \\
3.354\E12{$^\ddagger$\hspace{-10pt}} 
		  & 289.0 MeV & 3.699\E23 & 2.981\E23 & 2.451\E23 & 2.419\E23 & 6.650\E10 & 2.082\E19 & 5.027\E19 \\
2.000\E12 & 172.3 MeV & 6.204\E23 & 4.998\E23 & 4.124\E23 & 4.122\E23 & 8.820\E11 & 1.800\E20 & 1.831\E20 \\
1.000\E12 & 86.17 MeV & 1.241\E24 & 9.997\E23 & 8.381\E23 & 8.381\E23 & 2.821\E13 & 1.898\E21 & 1.898\E21 \\
5.000\E11 & 43.09 MeV & 2.481\E24 & 1.999\E24 & 1.769\E24 & 1.769\E24 & 9.017\E14 & 6.422\E21 & 6.422\E21 \\
4.597\E11{$^\dagger$\hspace{-10pt}} 
		  & 39.61 MeV & 2.699\E24 & 2.175\E24 & 1.944\E24 &	-		  & 1.372\E15 & 6.544\E21 & -         \\
2.000\E11 & 17.23 MeV & 6.204\E24 & 4.998\E24 & 4.925\E24 & -         & 8.744\E16 & 4.251\E20 & -         \\
1.069\E11{$^\ddagger$\hspace{-10pt}} 
		  & 9.212 MeV & 1.161\E25 & 9.352\E24 & 9.350\E24 & - 	 	  & 1.972\E18 & 3.145\E18 & -         \\
1.000\E11 & 8.617 MeV & 1.241\E25 & 9.997\E24 & 9.996\E24 & -         & 2.745\E18 & 3.418\E18 & -         \\
5.000\E10 & 4.309 MeV & 2.481\E25 & 2.000\E25 & 2.000\E25 & -         & 8.300\E19 & 8.300\E19 & -         \\ 
2.000\E10 & 1.723 MeV & 6.204\E25 & 5.004\E25 & 5.004\E25 & -         & 6.372\E21 & 6.372\E21 & -         \\
1.000\E10 & 861.7 keV & 1.241\E26 & 1.005\E26 & -         & -         & 1.263\E23 &	-	      & -         \\
\rowfont{\color{red}}
5.000\E9  & 430.9 keV & 2.481\E26 & 2.042\E26 & -         & -         & 1.514\E24 & -         & -         \\
\rowfont{\color{red}}
2.000\E9  & 172.3 keV & 6.204\E26 & 5.564\E26 & -         & -         & 9.795\E24 & -         & -         \\
\rowfont{\color{red}}
1.698\E9{$^\dagger$\hspace{-10pt}}  
		  & 146.3 keV & 7.307\E26 & 6.716\E26 &	-		  &	-		  & 1.037\E25 & -         & -         \\
\rowfont{\color{red}}
1.000\E9  & 86.17 keV & 1.241\E27 & 1.217\E27 & -         & -         & 5.107\E24 & -         & -         \\
\rowfont{\color{red}}
5.000\E8  & 43.09 keV & 2.481\E27 & 2.481\E27 & -         & -         & 1.063\E23 & -         & -         \\ 
\rowfont{\color{red}}
2.000\E8  & 17.23 keV & 6.204\E27 & 6.204\E27 & -         & -         & 3.120\E16 & -         & -         \\
\end{tabu}
\end{center}
\end{small}
\caption{Numerical values for shear and bulk viscosities for 7, 9, 11 and 13 particles. The local maximas and minimas in the bulk viscosities are marked by $\dagger$ and $\ddagger$, respectively. Everything after neutrinos decoupling should be considering non-physical and are marked in a red color. At temperatures of $10^{14}$ K and above, the viscosity due to electromagnetic interacting particles should be considered (shown in orange).}
\label{Tab:ViscosityTableSmall}
\end{table*}

\section{Conclusions}
The model universe we have presented here is a very simplified version of how the universe actually were at temperatures above $10^{12}$ K $= 100$ MeV. Our aim for this paper is to give a more illustrative description of early universe viscosity, particularly how it behaves with a multiple of different particles.

One important event which should be included however is the reheating effect during particle annihilations. Another interesting thought is whether we can use the same transport equations to get a better understanding of the neutrino decoupling? 

This extended model could be used to include new exotic particles, like dark matter candidates, or as a basis for an extended model using strongly interacting particles (e.g. pions) as well.

\acknowledgments
Thanks to my supervisor K\aa re Olaussen for giving me the project, and for helping me  throughout this work. His help has been invaluable for understanding  the concepts and  the underlying mathematics.  Thanks also  to Iver Brevik for introducing us to the field of viscous cosmology, and for fruitful discussions on this paper.


\appendix

\onecolumn

\section{Coefficients occurring in the linear equations}
\label{App:Coefficients}
We begin be rewriting the equations for shear and bulk viscosities 
\begin{align}
  \eta_s &= \frac{1}{10} cn \left( \frac{\kB T}{c} \right)^3 \sum_k c_k \gamma_k,
  \label{Eq:EtaSAppendix}\\  
  \eta_v &= n\kB T \sum_k a_k \alpha_k,
  \label{Eq:EtaVAppendix}
\end{align}
The $c_k$ and $a_k$ coefficients are solutions to the linearized relativistic Boltzmann (transport) equation for the shear and bulk cases
are given as follows:
\begin{equation}
c_k = \frac{1}{n} \left(\frac{\kB T}{c} \right)^2 c_{kl}^{-1} \gamma_k,
\end{equation}
\begin{equation}
a_k = \frac{1}{n} a_{kl}^{-1} \alpha_k,
\end{equation}
where
\begin{equation}
c_{kl}=x_k \left( x_l c'_{kl} + \delta_{kl} \sum_{m=1}^N x_m c''_{km} \right),
\end{equation}
\begin{equation}
a_{kl}=x_k \left( x_l a'_{kl} + \delta_{kl} \sum_{m=1}^N x_m a''_{km} \right)
\end{equation}
are matrix elements related to the differential cross-sections in the frame of the center of momentum of the system. The primes and double primes are defined as
\begin{align}
c'_{kl}&=\frac{1}{3}\left(\frac{\kB T}{c}\right)^4 \left(-40I^1_{12110}-80I^1_{13110}+6I^2_{21000}+12I^2_{22000}+8I^2_{23000}\right),
\\
c''_{kl}&=\frac{1}{3}\left(\frac{\kB T}{c}\right)^4 \left(40I^1_{12200}+80I^1_{13200}+6I^2_{21000}+12I^2_{22000}+8I^2_{23000}\right),
\\
a'_{kl} &= - 2I_{12000},
\\
a''_{kl} &= 2I_{12000} = -a'_{kl}.
\end{align}
Here, as in the Hoogeveen article, the particle indexes $k$ and $l$ are suppressed on the right sides of the equations. The 10+2 I-integrals described above are collision integrals, and have dimension $\sigma c$ [cm$^3/$s] and are defined by \citep{vanErkelens:RelativisticBoltzmannTheoryV}:
\begin{equation}
\begin{aligned}
  I^{\mu}_{i j \kappa \lambda h}(k,l) = 
  & \left( \frac{c\pi}{z_k^2z_l^2 K_2(z_k) K_2(z_l)} \right) \left( \frac{c}{\kB T} \right)^{2i-j+\kappa+\lambda+6} \\
  &\times \int_{(m_k+m_l)c}^\infty dP_{kl} g_{kl}^{2i+2} P_{kl}^{-j+3}(g^2_{kl}+m_k^2c^2)^{\kappa/2}(g^2_{kl}+m_l^2c^2)^{\lambda/2} \\
  &\times K_{j+h} \left( \frac{cP_{kl}}{\kB T} \right) \int_0^\pi (1-\cos^\mu \theta_{kl}) \frac{d\sigma_{kl}}{d\Omega_{\text{CM}}} (P_{kl},\theta_{kl} ) \sin \theta_{kl} d\theta_{kl},
\end{aligned}
\label{Eq:IIntegral}
\end{equation}
where
\begin{equation}
  g_{kl}=P_{kl}^{-1} \sqrt{ \tfrac{1}{4}(P_{kl}^2-m_k^2 c^2-m_l^2 c^2)^2 - m_k^2m_l^2c^4 }
  \label{Eq:g}
\end{equation}
is the invariant relative momentum.

\section{Weak interaction cross-sections}
\label{App:CrossSections}
The weak interaction cross-sections are given in Appendix C in \citep{Hoogeveen:1986}. However we have included the tau particles in our model to give us the differential cross-sections for 68 (compared to 45) elastic collisions
\begin{equation}
  k+l \rightarrow k+l,
  \label{Eq:ElasticKL}
\end{equation}
where
\begin{align}
  k &= \nu_e, \overline{\nu}_e, \nu_\mu, \overline{\nu}_\mu, \nu_\tau, \overline{\nu}_\tau,
  \label{Eq:KParticles}\\
  l &= \nu_e, \overline{\nu}_e, \nu_\mu, \overline{\nu}_\mu, \nu_\tau, \overline{\nu}_\tau, e^-, e^+, \mu^-, \mu^+, \tau^-, \tau^+.
  \label{Eq:LParticles}
\end{align}
The weak interaction cross-sections from Eq. (\ref{Eq:IIntegral}) is given as:
\begin{equation}
\frac{d\sigma_{kl}}{d\Omega_{\text{CM}}} = \frac{\GF^2 s}{32\pi^2 \hbar^4c^2} Sf(s,t),
  \label{Eq:WeakCrossSection}
\end{equation}
where $S$ is equal to the product of the usual symmetry factor $1/n_\text{f}!$, where $n_\text{f}$ is the number of identical particles in the final state, and a factor 2 for each incoming neutrino, so e.g. for $e^+e^+ \rightarrow e^+e^+$ we have $S=(1\times1)/2!$ and for $\nu_e\nu_e \rightarrow \nu_e\nu_e$ we have $S=(2\times2)/2!$. $f(s,t)$ is a collisions function for the two incoming particles and is given by:
\begin{equation}
f(s,t)=(v+a)^2\left(\frac{s-m^2c^2}{s}\right)^2+(v-a)^2\left(\frac{s+t-m^2c^2}{s}\right)^2+(v^2-a^2)\frac{2m^2c^2t}{s^2},
\end{equation}
with the $a$ and $v$ matrices are related to the neutral current coupling \citep{Hoogeveen:1986}, basically how neutrinos couples to the other particles. They are listen in Table \ref{Tab:va}. The $s$ and $t$ being the Mandelstam variables:
\begin{equation}
s \equiv (p_k+p_l)^2 = P^2_{kl},	
\label{Eq:s}
\end{equation}
\begin{equation}
t \equiv (p_k-p'_k)^2,	
\label{Eq:}
\end{equation}
where $s$ is the square of the center-of-mass energy (invariant mass) for the two incoming particles $k$ and $l$.
$t$ is the square of the four-momentum transfer, and is related to the scattering angle in the center of momentum system, being defined below. $p'_k$ is the four-momentum of particle $k$ after collision. 
such that
\begin{equation}
\begin{aligned}
t =&  m_k^2c^2+m_l^2c^2-P_{kl}^2+\frac{(P_{kl}^2+m_k^2c^2-m_l^2c^2)(P_{kl}^2-m_k^2c^2+m_l^2c^2)}{2P_{kl}^2}\\
   &+ \frac{[P_{kl}^2-(m_kc+m_lc)^2][P_{kl}^2-(m_kc-m_lc)^2]}{2P_{kl}^2}\cos \theta_{kl},
\end{aligned}
\end{equation}
with $\theta_{kl}$ being the scattering angle of the CM-system.

\begin{table}[h]
 \centering
 \subtable[$a$ matrix]
   {\includegraphics[width=0.48\textwidth]{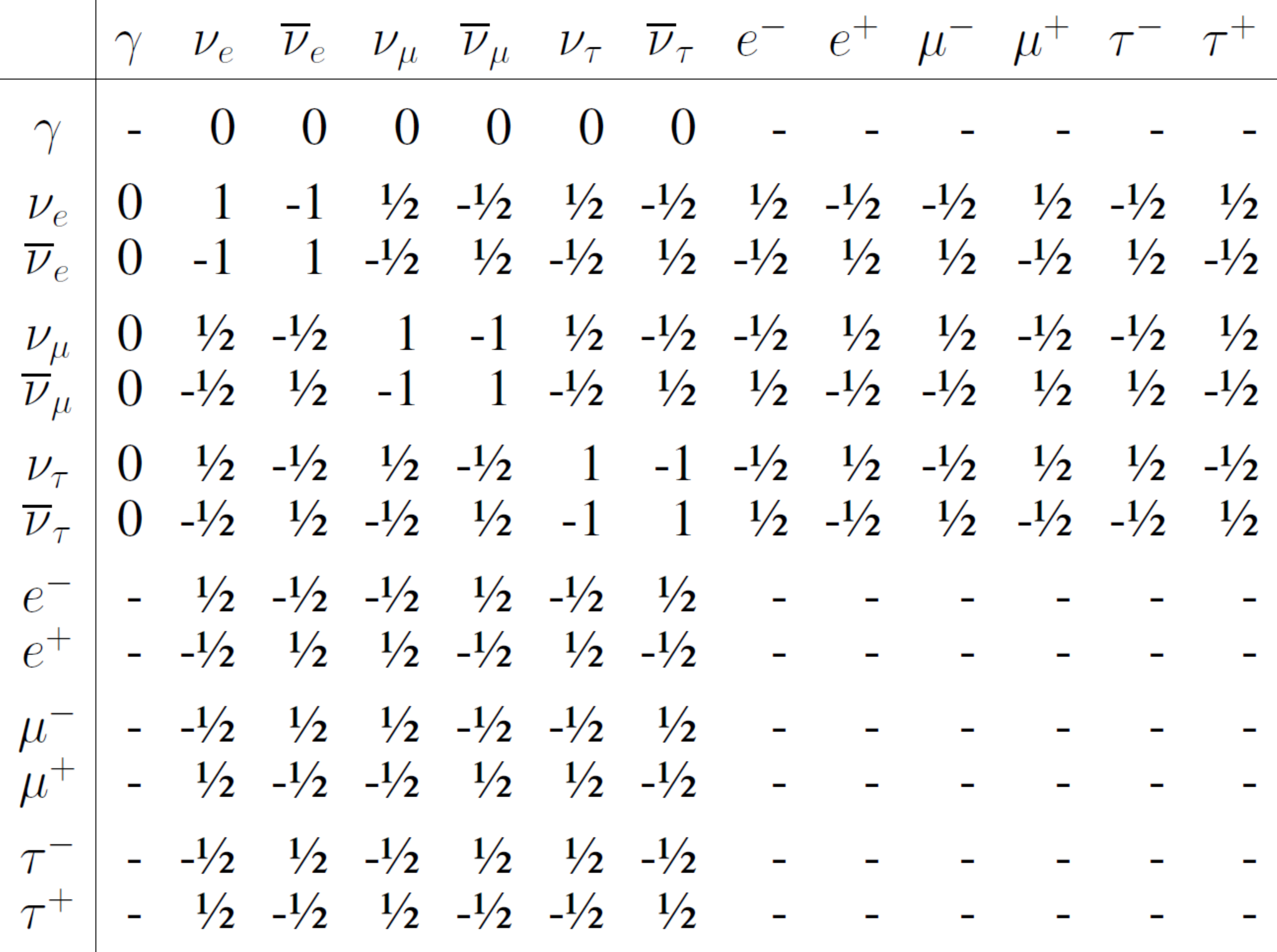}
   \label{TabSub:V} }
   \hspace{1mm}
 \subtable[$v$ matrix]
   {\includegraphics[width=0.48\textwidth]{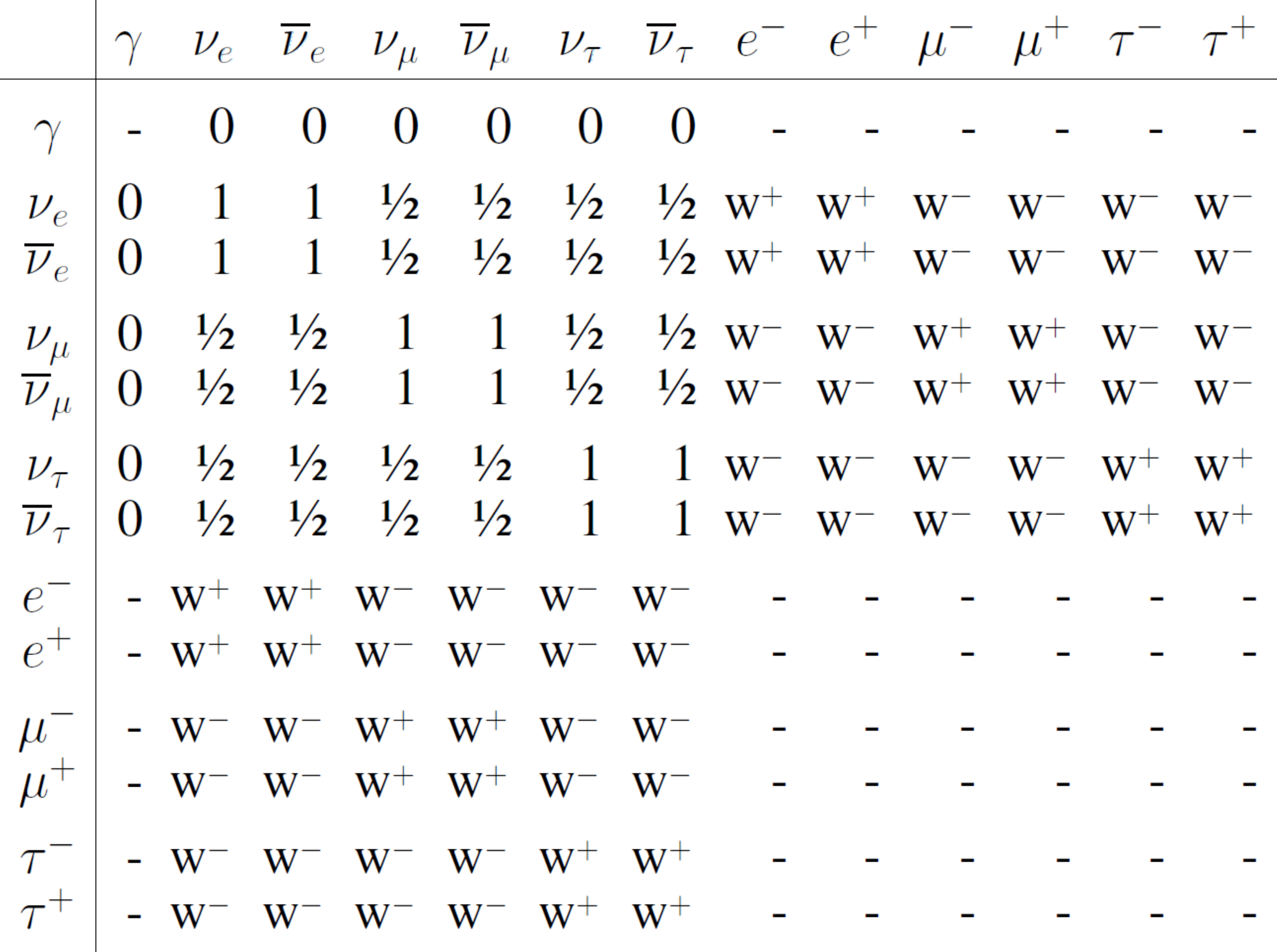}
   \label{TabSub:A} }
 \caption{$v$ and $a$ matrices. w$^+$ and w$^-$ is abbreviation for \W +\2 and \W -\2, respectively. The electromagnetic contribution is so small it can be ignored. The crossection $\sigma_{\gamma \nu}$ is zero.}
 \label{Tab:va}
\end{table}

\bibliographystyle{spr-mp-nameyear-cnd}
\bibliography{Bibliography}

\end{document}